%% file: mypaper.tex
\pdfoutput=1

\documentclass[11pt]{article}

\usepackage[final]{acl}
\usepackage{times}
\usepackage{latexsym}
\usepackage{algorithm}
\usepackage{algorithmicx}
\usepackage{amsfonts,amssymb,amsmath}
\usepackage{algpseudocode}
\usepackage{amsmath}
\usepackage{bbm}
\usepackage[T1]{fontenc}

\usepackage[utf8]{inputenc}
\usepackage{graphicx}
\usepackage{multirow}
\usepackage{booktabs}
\usepackage{graphicx}  
\usepackage{booktabs}
\usepackage{microtype}

\usepackage{inconsolata}
\usepackage[many]{tcolorbox}
\usepackage{graphicx}
\usepackage{xspace}
\newcommand{\themodel}{SINCon\xspace}
\title{SINCon: Mitigate LLM-Generated Malicious Message Injection Attack for Rumor Detection}

\nocopyright

\author{Mingqing Zhang\textsuperscript{1,2}, \ Qiang Liu\textsuperscript{1,2}, \ Xiang Tao\textsuperscript{1,2}, \ Shu Wu\textsuperscript{1,2},  \ Liang Wang\textsuperscript{1,2}\\
  \textsuperscript{\rm 1}New Laboratory of Pattern Recognition (NLPR)\\
State Key Laboratory of Multimodal Artificial Intelligence Systems\\
Institute of Automation, Chinese Academy of Sciences\\
    \textsuperscript{\rm 2}School of Artificial Intelligence, University of Chinese Academy of Sciences\\
    mingqing.zhang@cripac.ia.ac.cn,
    xiang.tao@cripac.ia.ac.cn, \{qiang.liu, shu.wu, wangliang\}@nlpr.ia.ac.cn
  }

\begin{document}
\maketitle
\begin{abstract}
In the era of rapidly evolving large language models (LLMs), state-of-the-art rumor detection systems, particularly those based on Message Propagation Trees (MPTs), which represent a conversation tree with the post as its root and the replies as its descendants, are facing increasing threats from adversarial attacks that leverage LLMs to generate and inject malicious messages. Existing methods are based on the assumption that different nodes exhibit varying degrees of influence on predictions. They define nodes with high predictive influence as important nodes and target them for attacks. 
If the model treats nodes' predictive influence more uniformly, attackers will find it harder to target high predictive influence nodes.
In this paper, we propose \textbf{S}imilarizing the predictive \textbf{I}nfluence of \textbf{N}odes with \textbf{Con}trastive Learning (\textbf{SINCon}), a defense mechanism that encourages the model to learn graph representations where nodes with varying importance have a more uniform influence on predictions. 
Extensive experiments on the Twitter and Weibo datasets demonstrate that \textbf{SINCon} not only preserves high classification accuracy on clean data but also significantly enhances resistance against LLM-driven message injection attacks.

\end{abstract}

\input{intro}

\input{relatedwork}

\input{preliminaries}

\input{method}
\input{experiment}
\input{conclusion}
\input{limitation}

\bibliography{references}

\appendix

\input{appendix}
\label{sec:appendix}

\end{document}

%% file: intro.tex
\section{Introduction}

\begin{figure}[t]
\setlength{\abovecaptionskip}{0pt}  
\setlength{\belowcaptionskip}{0pt}  
   \begin{center}
  \vspace{-4mm}
\includegraphics[width=\columnwidth]{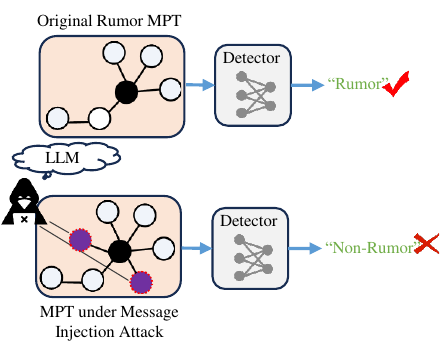}

   \end{center}
   \caption{The rumor detection model is attacked \\by LLM-generated malicious message injection. The message injection attack, generated by an LLM, introduces new nodes and edges, altering the topology and semantics of the MPT. This causes the rumor detection model to fail in effectively detecting the rumor.} 
   \label{fig:intro}
\end{figure}

The rapid advancement of large language models (LLMs) has revolutionized natural language processing, enabling impressive capabilities in text generation~\cite{li2024pre,huang2024grounded}, summarization~\cite{zhu2023calypso,xu2024mental}, and contextual reasoning~\cite{deng2024enhancing,kwon2024large}. However, these advancements also introduce new security challenges~\cite{zhan2023similarizing}, particularly in the domain of rumor detection on social media. 

Recent studies have revealed that Message Propagation Trees (MPTs), modeled as conversation trees with the root representing the source post and subsequent nodes representing retweets or comments, are vulnerable to malicious message injection attacks when used in rumor detection models that leverage graph neural networks (GNNs) to analyze message propagation patterns~\cite{liu2018early,zhang2019adversarial,song2021adversary}. Attackers can exploit LLMs to generate and inject deceptive messages into MPTs, significantly altering their topological and semantic structure. 
As a result, even state-of-the-art rumor detection models can be misled into classifying rumors as non-rumors, undermining their effectiveness in mitigating misinformation~\cite{sun2024exploring,li2025semantic}. As shown in Figure\ref{fig:intro}, the attacker leverages LLM to conduct a message injection attack on MPTs, successfully bypassing the rumor detector.

Previous methods for attacking MPT-based rumor detection models rely on the assumption: \textbf{different nodes in an MPT contribute unequally to the model's prediction, with important nodes having a greater predictive influence than unimportant ones}~\cite{mkadry2017towards,zou2021tdgia,luo2024message}. Therefore, based on the node importance scores obtained through attribution approaches, the attack can be viewed as an iterative process where the most important nodes are targeted first. Consequently, the imbalance in predictive influences of nodes within the MPT leads to a critical vulnerability. Attackers can design targeted attacks that focus on high-influence nodes, triggering a chain reaction that disrupts the overall propagation structure. 

Building on the aforementioned assumption, the success of attacks against MPT-based rumor detection models becomes clear. In a MPT, nodes can be categorized into important nodes, which carry more influential information for prediction, and unimportant nodes, which contribute less. Attack methods that target important nodes first are able to perturb the most critical information at each step, thereby making the model more susceptible to deception. Therefore, a counterintuitive question naturally arises: \textbf{Would the model be more robust if both important and unimportant nodes exerted a similar degree of influence on its predictions? }

To explore the aforementioned question, We propose \textbf{S}imilarizing the predictive \textbf{I}nfluence of \textbf{N}odes with \textbf{Con}trastive Learning (\textbf{SINCon}), a self-supervised regularization method designed to enhance model robustness against adversarial message injection attacks. SINCon mitigates the effect of localized perturbations by ensuring that both high- and low-influence nodes contribute more evenly to model predictions. Specifically, we define important and unimportant nodes as the top and bottom 10\% of nodes ranked by influence scores within the MPT. To regularize the model, we introduce two data augmentation strategies: one that masks important nodes and another that masks unimportant nodes.
 SINCon then leverages a contrastive learning objective to (1) reduce the disparity in model predictions between these two augmented MPTs, ensuring that nodes of different influence levels have a more uniform impact, (2) maintain similarity between the augmented MPTs and the original MPT, preventing excessive information loss, (3) minimize the agreement between the original MPT and other distinct MPTs within the same batch, avoiding the trivial solution of pattern collapse and encouraging the model to learn
more discriminative and robust representations.
 
We conduct extensive experiments on Twitter and Weibo datasets, evaluating SINCon against state-of-the-art MPT-based rumor detection models under LLM-generated malicious message injection attack. Our results demonstrate that by integrating SINCon into the training process, we effectively reduce the model's sensitivity to adversarial message injections, making it significantly more resilient to LLM-driven attacks while maintaining high performance on clean data.

Our main contributions can be summarized as follows:
\begin{itemize}
    \item We identify the imbalance in node influence within MPT-based rumor detection models, which makes them vulnerable to malicious message injection attacks.
    \item We introduce SINCon, a contrastive learning method that balances node influence, reducing the model’s vulnerability to attacks.
    \item Extensive experiments on Twitter and Weibo datasets show that SINCon improves model robustness to LLM-driven attacks while maintaining high performance on clean data.
\end{itemize}

%% file: relatedwork.tex
\section{Related Work}

\begin{figure*}[t]
\setlength{\abovecaptionskip}{0pt}  
\setlength{\belowcaptionskip}{0pt}  
   \begin{center}
\includegraphics[width=1\textwidth]{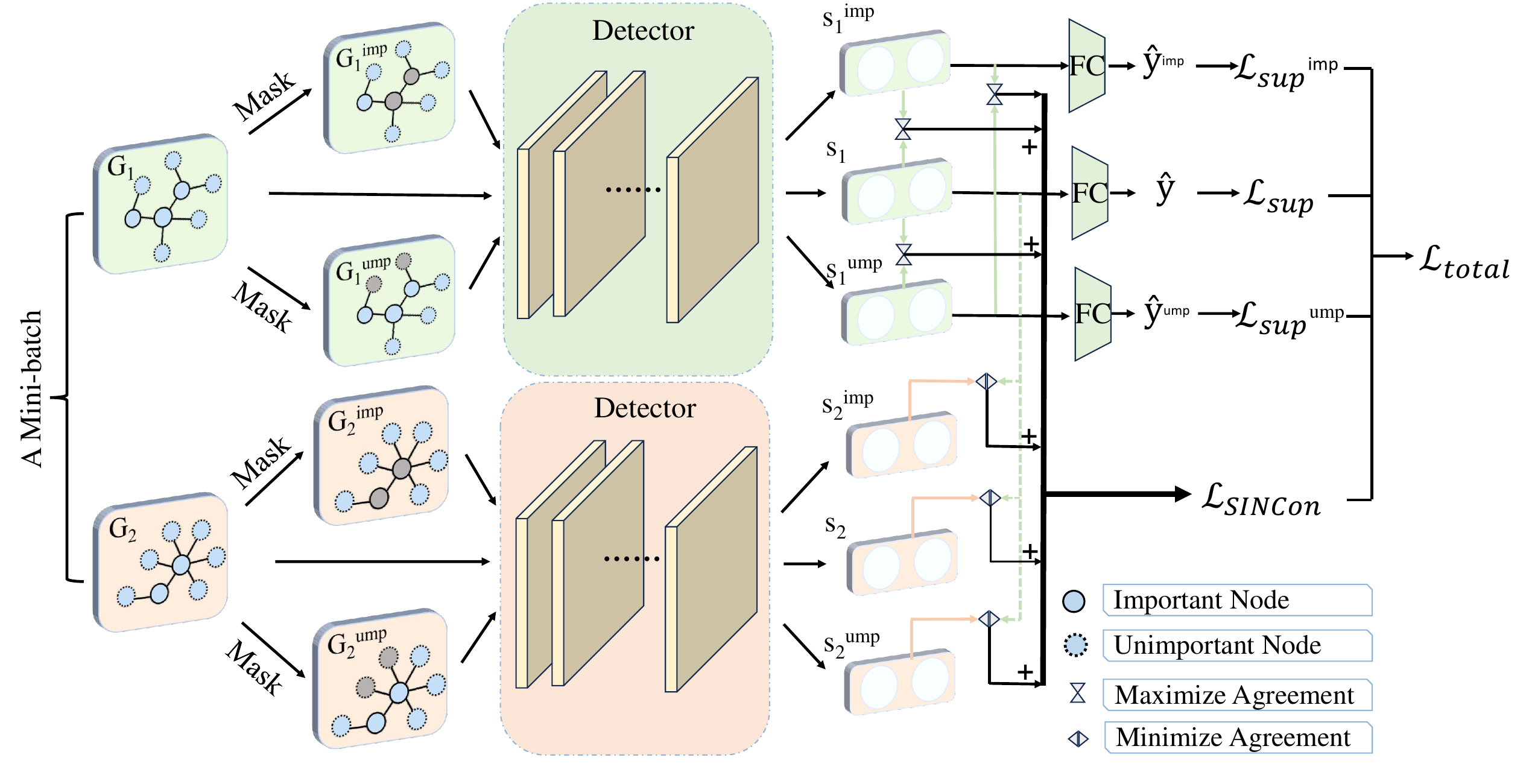}
   \end{center}
   \caption{Architecture of \themodel. Given a mini-batch \( G_i \in \{G_i\}_{i=1}^B \) of MPTs, where \( B = 2 \): (1) we define the top 10\% of nodes with the highest and lowest influence scores in an MPT as important and unimportant nodes, respectively, based on Eq. \ref{eq:12}. (2) To regularize the model, we introduce two data augmentation strategies: one that masks important nodes and another that masks unimportant nodes. (3) reduce the disparity in model predictions between these two augmented MPTs, maintain similarity between the augmented MPTs and the original MPT, minimize the agreement between the original MPT and other distinct MPTs within the same batch.  } 
   \label{fig:2}
\end{figure*}

\noindent\textbf{MPT-based Rumor Detection.} Rumor detection on social media aims to identify and prevent the spread of misinformation. Recent methods use GNNs to capture information from MPTs~\cite{wu2020rumor,xu2022evidence,zhang2023rumor,wu2023adversarial,tao2024semantic}. An MPT-based rumor detection model typically has three components: (1) message encoding, (2) GNN and (3) a readout function. Different studies use varied approaches for these components, such as word frequency counts~\cite{malhotra2020classification,khoo2020interpretable,sun2022rumor} or dense embeddings for encoding~\cite{liu2024rumor,tao2024out,zhang2024breaking}, and GCN or GAT for learning propagation patterns~\cite{wu2022bias,xu2022evidence,zhang2024t3rd}. The readout function often combines strategies like mean or max aggregation. In this paper, we mainly applied several state-of-the-art MPT-based rumor detection models to investigate their robustness.

\noindent\textbf{LLM-Generated Attacks.}
Large Language Models (LLMs) are capable of generating highly coherent and contextually relevant text~\cite{liu2023agentbench,yang2024harnessing,zhu2024llms,valmeekam2023planning}. However, while large language models demonstrate significant capabilities, they are increasingly drawing attention due to their generation of malicious information~\cite{kreps2022all}.
Recent research has shown that content generated by LLMs is often indistinguishable from content created by humans~\cite{zhao2023more,uchendu2023does}.Some works considered the use of LLMs for rumor generation~\cite{huang2022faking,lucas2023fighting,pan2023risk}. In this work, we aim to investigate methods for detecting such LLM-generated rumors and propose a defense mechanism to mitigate their impact on rumor detection systems.

\noindent\textbf{Adversarial Attacks and Defenses in Rumor Detection.} Rumor Detection Model adversarial attacks include evasion attacks~\cite{luo2024message} and poisoning~\cite{li2023pagcl}, as well as global~\cite{fang2024gani} and targeted types~\cite{zhang2023minimum}. With the rapid development of LLM technology, LLMs have become tools for attackers~\cite{hu2024llm,xu2023llm}. These attacks exploit the capabilities of LLMs to craft misleading messages or manipulate the structure of MPTs, resulting in subtle alterations to node features or edge relationships that deceive the model into making incorrect predictions. Adversarial samples are commonly used in various studies to train GNNs with enhanced robustness via adversarial training techniques~\cite{gosch2024adversarial,zhai2023state,zhang2023sam}. However, due to the scarcity of adversarial samples, the effectiveness of these methods is often limited. In this work, we introduce the technique of similarizing the influence of nodes with Contrastive Learning to enhance the robustness of rumor detection models.

%% file: preliminaries.tex
\newtcolorbox{mybox}{
    boxrule = 1.0pt,
    rounded corners,
    colback = sub,
    left = 1mm,
    right = 1mm,
    top = 1mm,
    bottom = 1mm,
    before skip = 1.6mm,
    arc = 0pt   
}
\definecolor{sub}{HTML}{eeeeee}

\section{Preliminaries}

\subsection{MPT-based Rumor Detection}

\noindent\textbf{Message Propagation Tree.} Let $\mathcal{G} = \{G_i\}_{i=1}^{|\mathcal{G}|}$ be a set of MPTs, where each MPT 
$G_i = (\mathcal{X}_i, \textbf{A}_i)$
consists of a set of messages 
$\mathcal{X}_i = \{x_1^{(i)}, x_2^{(i)}, \dots, x_{n_i}^{(i)}\}$,
and an adjacency matrix $\textbf{A}_i\in\{0,1\}^{n_i\times n_i}$ indicating reply or retweet relations. Here, $x_1^{(i)}$ is the source post and $\{ x_j^{(i)} \}_{j=2}^{n_i}$, representing comments, replies, or retweets related to the source post, $n_i$ denotes the number of messages in the $i$-th MPT. Each MPT has a binary label $y_i\in\{y_r,y_{nr}\}$, where $y_r$ and $y_{nr}$ represent  rumor and non-rumor classifications, respectively. 

We split the dataset into $\mathcal{G}_\text{train}$ and $\mathcal{G}_\text{test}$, corresponding to the training and testing sets of MPTs, respectively. The goal of MPT-based rumor detection is to train a binary classifier $f_{\theta}(G)$, parameterized by $\theta$, using $\mathcal{G}_\text{train}$. The classifier $f_\theta(G)$ is trained on the training set $\mathcal{G}_{\text{train}}$ by minimizing the loss:
\begin{equation}\label{eq:1}
\mathcal{L}_{\text{sup}}(f_\theta(G)) = \sum_{G_i \in \mathcal{G}_{\text{train}}} \mathcal{L}(f_\theta(G_i), y_i),
\end{equation}
with optimal parameters
\begin{equation}\label{eq:2}
\theta^* = \arg\min_\theta \mathcal{L}_{\text{sup}}(f_\theta(G)).
\end{equation}
The trained model predicts $\hat{y}_i = f_{\theta^*}(G_i)$ for unseen MPTs in $\mathcal{G}_{\text{test}}$. 

\noindent\textbf{MPT-based Rumor Detector.}
Messages are encoded into feature vectors using an encoding function $\mathcal{E}(\cdot)$:
\begin{equation}\label{eq:3}
\textbf{H}^{(0)} = \mathcal{E}(X) = [\textbf{h}_1^{(0)}, \textbf{h}_2^{(0)}, \dots, \textbf{h}_{n_i}^{(0)}].
\end{equation}
A GNN is then applied to learn both propagation patterns and content. For each message $x_u$, the feature update at layer $l$ is:
\begin{align}\label{eq:4}
\textbf{h}_u^{(l)} = & \sigma^{(l-1)} \left( \textbf{h}_u^{(l-1)}, \right. \nonumber \\
           & \left. \mathcal{AGG}_{x_v \in N(x_u)} \left( \gamma^{(l-1)}(\textbf{h}_v^{(l-1)}, \textbf{h}_u^{(l-1)}) \right) \right),
\end{align}
where $\sigma(\cdot)^{(l)}$ and $\gamma(\cdot)^{(l)}$ are the activation functions at the $l$-th layer of the GNN, $x_v \in N(x_u)$ is the 1-hop retweets or comments of message $x_u$, and $\mathcal{AGG}(\cdot)$ represents the aggregation operation.
A readout function $\mathcal{R}(\cdot)$ aggregates these features into a summary representation:
\begin{equation}\label{eq:5}
\textbf{s} = \mathcal{R}(\textbf{H}^{(L)}).
\end{equation}
Finally, the prediction is given by:
\begin{equation}\label{eq:6}
\hat{y} = \text{Softmax}(\textbf{s}).
\end{equation}

\subsection{Message Injection Attack}

\noindent\textbf{Objective of the Attack.} We denote $\mathcal{G}_r \subseteq \mathcal{G}_{\text{test}}$ and $\mathcal{G}_{nr} \subseteq \mathcal{G}_{\text{test}}$ 
as the set of rumor and non-rumor MPTs in the testing MPT set $\mathcal{G}_{\text{test}}$, respectively.
The goal of the attack is to deceive the rumor detector into misclassifying a rumor MPT \( G \in G_r \) as a non-rumor MPT by injecting a set of malicious messages \( \mathcal{X}_{\text{atk}} \) into the MPT, with the constraint \( | \mathcal{X}_{\text{atk}} | \leq \Delta \) and \( d_{\text{in}}(x_u) = 1, \forall x_u \in \mathcal{X}_{\text{atk}} \). The budget \( \Delta \) refers to the maximum number of malicious messages that can be injected into the MPT. This constraint ensures that the attack remains inconspicuous. The attacker minimizes the negative testing loss:
\begin{equation}\label{eq:7}
\min \sum_{G \in G_r} - \mathcal{L}_{\text{sup}}(f_\theta^*(G')),
\end{equation}
where \( G' = (\mathcal{X}', A') \) is the MPT with injected malicious messages.

\noindent\textbf{Message Pair and Root-Centric Homophily.} The attack effectiveness relies on disrupting the MPT homophily distribution. The message pair homophily between messages \( x_u \) and \( x_v \) is defined as:
\begin{equation}\label{eq:8}
\text{sim}(x_u, x_v) = \frac{\textbf{h}_u^{(0)} \cdot \textbf{h}_v^{(0)T}}{\| \textbf{h}_u^{(0)} \|_2 \| \textbf{h}_v^{(0)} \|_2}.
\end{equation}
The root-centric homophily measures the similarity between the source post \( x_1 \) and other messages in the MPT:
\begin{equation}\label{eq:9}
\text{sim}_{\text{root}}(G) = \frac{1}{n} \sum_{x_j \in \{x_j\}_{j=2}^n} \text{sim}(x_j, x_1).
\end{equation}

\noindent\textbf{Iterative Malicious Message Generation.} To generate malicious messages, we employ system prompt which is demonstrated in Appendix A.\ref{A.1}. The process begins with the system prompt \( p \) and the source post \( x_1 \), producing an initial malicious message :
\begin{equation}\label{eq:10}
 x_{\text{atk}} = \text{LLM}(x_1, p). 
\end{equation}
If the root-centric homophily of the generated message exceeds a threshold \( \lambda \), the prompt is refined iteratively:
\begin{equation}\label{eq:11}
x_{\text{atk}}' = \text{LLM}(x_{\text{atk}}, p'),
\end{equation}
where \( p' \) contains the homophily information, as detailed in Appendix A.2.

\noindent\textbf{Connecting Malicious Messages.}
The more
a message and its neighboring messages are commented on or
retweeted, the higher the influence that message holds in the final
summary representation \textbf{s}~\cite{luo2024message}.The generated malicious messages are connected to existing messages in the MPT based on their influence score, which is calculated as:
\begin{equation}\label{eq:12}
I_{x_u} = \sqrt{{d_{x_u}}{d_{x_v}}}, \quad x_v \in \mathcal{N}(x_u) \cup \{x_u\}.
\end{equation}
The malicious message is then connected to the message with the highest influence score, updating the adjacency matrix \( A \) to \( A' \).

\noindent\textbf{Attack Procedure.} For each rumor MPT \( G \in \mathcal{G}_{\text{test}} \), the LLM-based Message Injection Attack generates malicious messages and injects them into the MPT. If the prediction of the MPT is a non-rumor \( \hat{y} = \text{y}_{nr} \), the message injection stops for that MPT.

%% file: method.tex
\section{Method}

Recall that the goal of \themodel is to similarize the influence of nodes. To formally define this goal, we first define the 10\% of nodes in an MPT with the highest and lowest influence scores as the important and unimportant nodes, respectively, according to Eq.\ref{eq:12}.
 We then propose two data augmentation operations, $t^{imp}(\cdot)$ and $t^{ump}(\cdot)$, which respectively means mask important and unimportant nodes in a MPT. Therefore, under the training scenario of Eq.\ref{eq:1}, the primary goal of \themodel can now be formulated as:

\begin{flalign}\label{eq:13}
\min_{\theta} \|\mathcal{Q}_{\text{imp}} - \mathcal{Q}_{\text{ump}}\|: &&
\end{flalign}
\begin{align*}
\mathcal{Q}_{\text{imp}} &= \mathop{\mathbb{E}}\limits_{\substack{ G^{\text{imp}} \sim t^{\text{imp}}(G)}}
\Bigl[  \mathcal{P}( G) - \mathcal{P}(G^{\text{imp}}) \Bigr], \\[1ex]
\mathcal{Q}_{\text{ump}} &= \mathop{\mathbb{E}}\limits_{\substack{ G^{\text{ump}} \sim t^{\text{ump}}(G)}}
\Bigl[  \mathcal{P}( G) - \mathcal{P}(G^{\text{ump}}) \Bigr],
\end{align*}

Here, \(G^{\text{imp}}\) is an augmentation sampled from \(t^{\text{imp}}(G)\), and \(G^{\text{ump}}\) is an augmentation sampled from \(t^{\text{ump}}(G)\). \( \mathcal{P}(\cdot) \) represents the model's predicted probability distribution over the possible output classes for the input MPT. \(\mathcal{Q}^{\text{imp}}\) and \(\mathcal{Q}^{\text{ump}}\) measure the extent of model confidence decrease when information in the important and unimportant nodes is mask, indicating the overall influence of the information in nodes of different importance on prediction.

The complete objective of \themodel can be further decomposed into two perspectives:

\medskip

\textbf{Objective 1:} The influence of different nodes should be similar, thus the model should treat the MPT with information in nodes of different importance mask (\(G^{\text{imp}}\) and \(G^{\text{ump}}\)) similarly.

\medskip

\textbf{Objective 2:} The influence of different nodes should be slight, thus the model should treat the MPT with different information mask (\(G^{\text{imp}}\) and \(G^{\text{ump}}\)) similarly to the original MPT that contains complete information (\(G\)).

To achieve Objective 1 and Objective 2, and further the goal of \themodel, we use a contrastive loss objective from the perspective of MPT representation. To define the contrastive loss objective, for convenience, we first define the calculation \( S \):  
\begin{equation}\label{eq:14}
S^{(k,l)}_{(i,j)} = \exp\left(\text{sim}[\textbf{s}^k_i, \textbf{s}^l_j] / {\tau}\right),
\end{equation}
where \( k, l \in \{\text{imp}, \text{ump}, \cdot\} \), respectively indicate the augmentation sampled from \( t^{\text{imp}}(\cdot) \), the augmentation sampled from \( t^{\text{ump}}(\cdot) \), and the normal example. \( i, j \) are the example indices, \( \text{sim}[\textbf{r}_i, \textbf{r}_j] = \textbf{r}_i^\top \textbf{r}_j / {\|\textbf{r}_i\|\|\textbf{r}_j\|} \) is the cosine similarity, and \( \tau \) is a temperature parameter similar to the NT-Xent loss \citep{chen2020simple, oord2018representation}.  

Then the contrastive loss function for an example in a mini-batch \( G_i \in \{G_i\}_{i=1}^B \) is defined as:

\begin{equation}\label{eq:15}
\begin{split}
&\mathcal{L}_{\mathrm{SINCon}}(G_i; \theta) \\&= {} 
\mathop{\mathbbm{E}}\limits_{\substack{  G^{\text{imp}}_i \sim t^{\text{imp}}(G_i)\\G^{\text{ump}}_i \sim t^{\text{ump}}(G_i) }} 
\Biggl[ -\log \frac{S_{\mathrm{positive}}} {\sum_{j=1}^{B} S_{\mathrm{negative}} } \Biggr],
\end{split}
\end{equation}

where

\begin{equation}
\mathcal{S}_{\mathrm{positive}} = \mathcal{S}^{(\mathrm{imp},\mathrm{ump})}_{(i,i)} + \mathcal{S}^{(\cdot,\mathrm{ump})}_{(i,i)} + \mathcal{S}^{(\cdot,\mathrm{imp})}_{(i,i)}, 
\label{eq:16}
\end{equation}

\begin{equation}
\mathcal{S}_{\mathrm{negative}} = \mathcal{S}^{(\cdot,\cdot)}_{(i,j)} + \mathbbm{1}_{(i\neq j)} \cdot \Bigl[
\mathcal{S}^{(\cdot,\mathrm{ump})}_{(i,j)} + \mathcal{S}^{(\cdot,\mathrm{imp})}_{(i,j)}
\Bigr].
\label{eq:17}
\end{equation}

Let \( B \) be the batch size, and \( \mathbbm{1}_{(\cdot)} \) be an indicator function that equals 1 if the condition \( (\cdot) \) is true; otherwise, it equals 0. 
Specifically, to calculate the loss for each mini-batch, we first obtain the augmentations \( G^{\text{ump}}_i \) from \( t^{\text{ump}}(G_i) \) and the augmentations \( G^{\text{imp}}_i \) from \( t^{\text{imp}}(G_i) \) for each example in the mini-batch.

To achieve Objective 1, we use the term \( S^{(\text{imp},\text{ump})}_{(i,i)} \) in the numerator. This constraint maximizes the similarity between the representations of the augmentations with important and unimportant nodes removed, making the different degrees of incomplete information in the augmentations have a similar impact on the prediction.

To achieve Objective 2, we use the terms \( S^{(\cdot,\text{ump})}_{(i,i)} \) and \( S^{(\cdot,\text{imp})}_{(i,i)} \) in the numerator. These constraints maximize the similarity between the original MPT and the two augmentations, ensuring that the incomplete information in the remaining nodes of the augmentations has a similar influence as the complete information in the normal MPT.

Intuitively, the semantics of different examples should be distinct. Following the constraints in \( S_{\text{positive}} \), the semantics of the augmentations of different examples should also be different. Therefore, the three terms in \( S_{\text{negative}} \) indicate that, given an example within a mini-batch, both the other examples and the augmentations derived from other examples are treated as negative examples.

The final loss of \themodel regularization is computed across all examples in a mini-batch. When \themodel is used in the normal training scenario Eq.\ref{eq:1}, the overall objective is:
\begin{equation}\label{eq:18}
\begin{aligned}
&\min_{\theta} \mathcal{L}_{\text{total}}(\theta) \\&= {} \mathcal{L}_\text{sup}(f_{\theta}(G)) + \alpha_1(\mathcal{L}_\text{sup}(f_{\theta}(G^{imp})) \\
&+ \mathcal{L}_\text{sup}(f_{\theta}(G^{ump}))) + \alpha_2 \mathcal{L}_{\text{SINCon}}(G),
\end{aligned}
\end{equation}
where \( \alpha_1 \) and \( \alpha_2 \) are the parameters balancing the supervised part and the contrastive regularization part.

%% file: experiment.tex
\section{Experiment}

\begin{table*}[ht]
  \centering
  
    \begin{tabular}{ccccccc}
    \toprule
    \multicolumn{1}{c}{\multirow{2}[4]{*}{Surrogate Model}} & \multicolumn{1}{c}{\multirow{2}[4]{*}{Target Model}} & \multirow{2}[4]{*}{Method} & \multicolumn{2}{c}{Twitter} & \multicolumn{2}{c}{Weibo} \\
\cmidrule{4-7}          &       &       &   AUA.& ACC.  &AUA.   & ACC. \\
    \midrule
    \multirow{6}[6]{*}{BiGCN} & \multirow{2}[2]{*}{BiGCN} & Normal & 0.7604 & 0.8979 & 0.6457 & \textbf{0.9137} \\
          &       & w/ \themodel &\textbf{0.8833}  &\textbf{0.9021}  & \textbf{0.8697} & 0.9089 \\
\cmidrule{2-7}          & \multirow{2}[2]{*}{GACL} & Normal & 0.6250   & \textbf{0.9000} & 0.7325 & 0.8999 \\
          &       & w/ \themodel & \textbf{0.8458} & 0.8750 & \textbf{0.8930} &  0.8999\\
\cmidrule{2-7}          & \multirow{2}[2]{*}{GARD} & Normal & 0.6417 & \textbf{0.8854} &0.9096  & \textbf{0.9258} \\
          &       & w/ \themodel & \textbf{0.7750} & 0.8729 & \textbf{0.9237} & 0.9232 \\
    \midrule
    \multirow{6}[6]{*}{GACL} & \multirow{2}[2]{*}{BiGCN} & Normal & 0.8417 & \textbf{0.8979} & 0.5779 & \textbf{0.9153} \\
          &       & w/ \themodel & \textbf{0.8729} & 0.8708 & \textbf{0.7865} & 0.8898 \\
\cmidrule{2-7}          & \multirow{2}[2]{*}{GACL} & Normal &0.5354  & \textbf{0.8750} &   0.4931& \textbf{0.9200} \\
          &       & w/ \themodel & \textbf{0.8250} & 0.8583 & \textbf{0.8543} &  0.9041\\
\cmidrule{2-7}          & \multirow{2}[2]{*}{GARD} & Normal & 0.6604 & \textbf{0.8917} &0.9015  & \textbf{0.9359} \\
          &       & w/ \themodel &\textbf{0.8333}  & 0.8750 & \textbf{0.9174} & 0.9258 \\
    \midrule
    \multirow{6}[6]{*}{GARD} & \multirow{2}[2]{*}{BiGCN} & Normal & 0.7792   & \textbf{0.9000} & 0.5646 &\textbf{0.9110}  \\
          &       & w/ \themodel & \textbf{0.8333} & 0.8917 & \textbf{0.7969} & 0.8898 \\
\cmidrule{2-7}          & \multirow{2}[2]{*}{GACL} & Normal & 0.5667 & \textbf{0.9042} & 0.4995  & \textbf{0.9142}\\
          &       & w/ \themodel & \textbf{0.8500} & 0.8875  & \textbf{0.7797} & 0.8898 \\
\cmidrule{2-7}          & \multirow{2}[2]{*}{GARD} & Normal &0.6854  & \textbf{0.8917} &0.7188  & \textbf{0.9306} \\
          &       & w/ \themodel & \textbf{0.7708} & 0.8792 & \textbf{0.8231} & 0.9168 \\
    \bottomrule
    \end{tabular}%
    \caption{We compare model accuracy under attack (AUA.) and accuracy (ACC.). The \textbf{bold} values of AUA. and ACC. represent the strongest robustness and the highest accuracy, respectively. Normal refers to the standard rumor detection model, while w/ \themodel denotes the rumor detection model enhanced with \themodel.}
  \label{tab:1}%
\end{table*}%

\begin{table*}[htbp]
  \centering
  
    \begin{tabular}{ccccccc}
    \toprule
    \multicolumn{1}{c}{\multirow{2}[4]{*}{Surrogate Model}} & \multicolumn{1}{c}{\multirow{2}[4]{*}{Target Model}} & \multirow{2}[4]{*}{Method} & \multicolumn{2}{c}{Twitter} & \multicolumn{2}{c}{Weibo} \\
\cmidrule{4-7}          &       &       & AUA.  & ACC.  &AUA.   & ACC. \\
    \midrule
    \multirow{4}[4]{*}{BiGCN} & \multirow{2}[2]{*}{BiGCN} & SWICon &\textbf{0.8833}  & \textbf{0.9021} &\textbf{0.8697}  & \textbf{0.9089} \\
          &       & w/ random & 0.8178 & 0.8204 &0.8019  & 0.8427 \\
\cmidrule{2-7}          & \multirow{2}[2]{*}{GACL} & SWICon &\textbf{0.8458}  & \textbf{0.8750} &  \textbf{0.8930}&\textbf{0.8999}  \\
          &       & w/ random &0.8146  & 0.7646 & 0.8120 & 0.8056 \\
    \bottomrule
    \end{tabular}%
    \caption{Experimental results of \themodel with different data augmentation operation. w/ random means the augmentations of each MPT are sampled randomly rather than based on attributions.}
  \label{tab:2}%
\end{table*}%

\subsection{Datasets}
We use two real-world rumor datasets, Twitter~\cite{ma2017detect} and Weibo~\cite{ma2016detecting}, to evaluate the \themodel approach. These datasets are sourced from two popular social media platforms—Twitter and Weibo. The Twitter dataset consists of English rumor datasets with conversation threads in tweets, providing a rich context for analysis. On the other hand, the Weibo dataset comprises Chinese rumor datasets with a similar composition structure. These datasets are annotated with two labels: Rumor and Non-Rumor, which are used for the binary classification of rumors and non-rumors.  Detailed statistics for both
datasets are provided in Appendix A.\ref{tab:3}.

We employ two metrics to validate the effectiveness of the proposed method: accuracy under attack (AUA.) and Accuracy (ACC.). Note that the higher
the AUA. is, the more successful the defense method is. In contrast, a
low ACC indicates a reduced performance of the rumor detector after the attack. Our primary goal is to evaluate the performance of \themodel on both clean data and data subjected to Message Injection Attacks. therefore, we take the ACC and AUA. as our primary metrics.

\subsection{Settings}
In our preliminary experiments, we employed the state-of-the-art Message injection attack, i.e., HMIA-LLM ~\cite{luo2024message}, to attack four MPT-based state-of-the-art rumor detectors:
\begin{itemize}                     
\item\textbf{BiGCN}~\cite{bian2020rumor}: A GNN-based rumor detection model utilizing the Bi-directional propagation structure.
\item\textbf{GACL}~\cite{sun2022rumor}: A GNN-based model using adversarial and contrastive learning, which can not only encode the global propagation structure, but also resist noise and adversarial samples, and captures the event invariant features by utilizing contrastive learning.
\item \textbf{GARD}~\cite{tao2024semantic}: A rumor detection model introduces self-supervised semantic evolvement learning to facilitate the acquisition of more transferable and robust representations.
\end{itemize}

We simulated two attack scenarios for defense: one where the attacker uses the same model for both generating the attack content and launching the attack, and another where the attacker employs a surrogate model to generate the attack content, then to attack the target model (i.e., the model generating the attack content is not necessarily the same as the target model). We conducted experiments on both the standard rumor detection model (Normal) and the model enhanced with \themodel (w/ SINCon) to evaluate ACC. and AUA..

In executing HMIA-LLM, We followed the settings from the original study~\cite{luo2024message}, employing ChatGPT (gpt-3.5-turbo) to generate malicious messages, with a root-centric homophily threshold \( \lambda \) = 0.35 and a budget of \( \Delta \) = 50. 
Our experiments were conducted on a remote machine server with 1 NVIDIA RTX 3090 (24G) GPU. We set \( \alpha_1 \)=1e-5, \( \alpha_2 \)=1e-2 for Twitter, and \( \alpha_1 \)=1e-4, \( \alpha_2 \)=1e-4 for Weibo.

\subsection{Overall Performance}

The experimental results Table\ref{tab:1} shows that SINCon significantly enhances the robustness of MPT-based rumor detection models against LLM-generated message injection attacks. 

As shown in Table\ref{tab:1}, when combined with other rumor detection models, SINCon only introduces a slight negative effect on the accuracy of clean (Normal) data. Our analysis indicates that the maximum decrease in accuracy is 2.55\%, with some cases showing no decrease at all, and an average decline of 1.38\%. Overall, SINCon results in a modest reduction in model accuracy on clean data, a drop that primarily stems from the introduction of augmented samples used to implement the contrastive learning regularization.

SINCon significantly enhances the robustness of the model. As shown in Table \ref{tab:1}, across various rumor detection models, SINCon markedly improves performance when facing LLM-driven malicious message injection attacks, enabling the model to better resist adversarial perturbations. The analysis demonstrates that AUA achieves a maximum improvement of 36.12\%, a minimum improvement of 1.59\%, and an average improvement of 16.63\%. 
This approach substantially strengthens the model’s robustness in adversarial environments while maintaining high accuracy on clean data. Overall, the experimental results in Table \ref{tab:1} clearly demonstrate the exceptional effectiveness of SINCon in enhancing the model's resilience against LLM-generated malicious message injection attacks.

\subsection{Ablation Study}
\subsubsection{Data Augmentation Operation}
We conducted an ablation study to further explore the impact of the "Similarizing the Influence of Nodes" data augmentation operation on SINCon. Specifically, we replaced the data augmentation operations$t^{imp}(\cdot)$ and $t^{ump}(\cdot)$ in SINCon with a new data augmentation strategy that randomly masks nodes. This experiment was carried out using two different model combinations on both the Twitter and Weibo datasets. As shown in Table \ref{tab:2}, the performance (ACC. and AUA.) of the influence-based data augmentation strategy significantly outperforms the random node masking approach on both datasets. These results provide additional evidence that the method of similarizing the influence of nodes plays a crucial role in enhancing the robustness of SINCon in rumor detection models, effectively counteracting the impact of adversarial attacks. This finding further solidifies our hypothesis that balancing node influence improves the model's overall performance and resilience.

\begin{figure}
    \centering
    \includegraphics[width=0.9\linewidth]{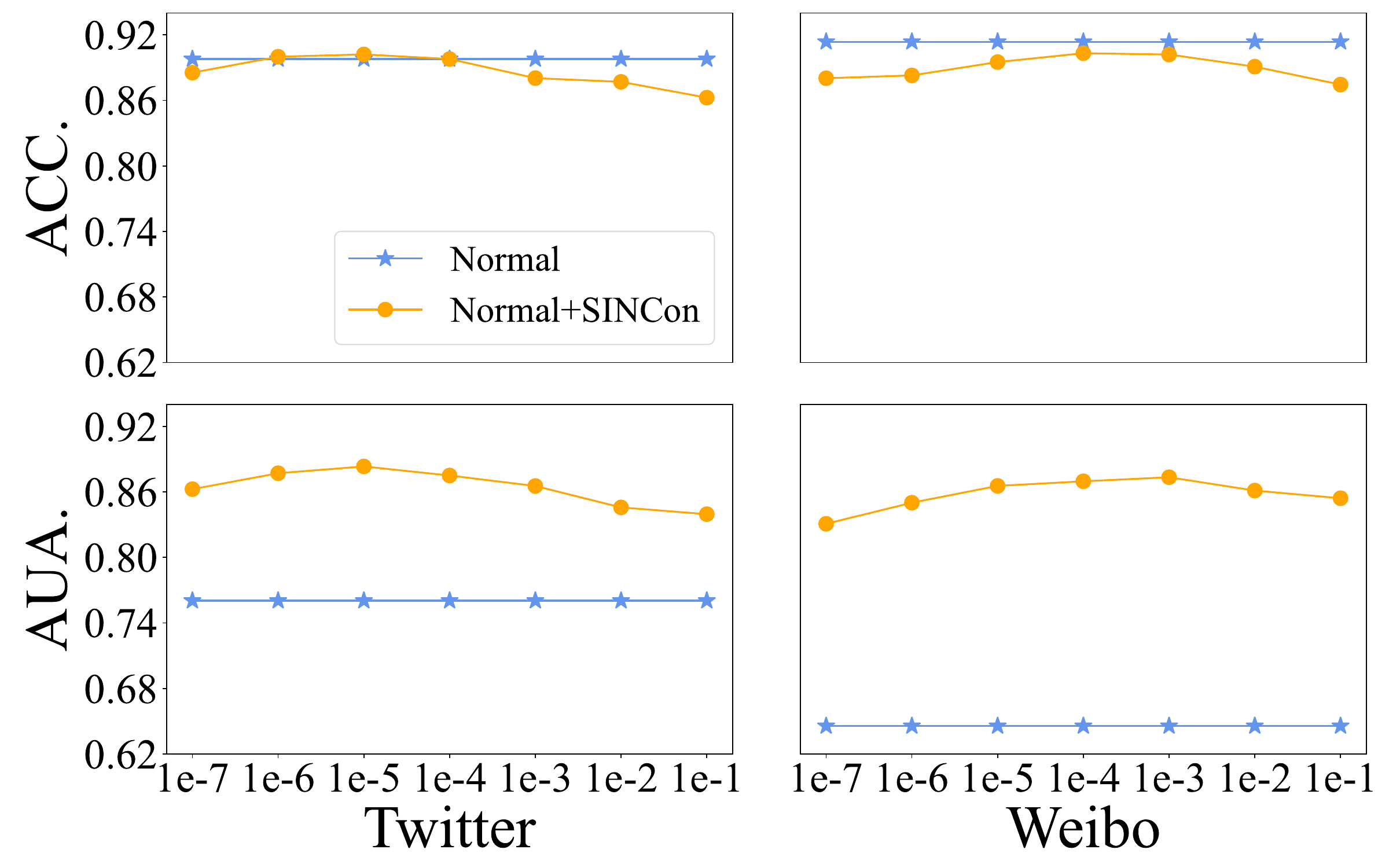}
    \vspace{-0.3cm}
    \caption{Sensitivity analysis of hyperparameters \(\alpha_1\). Experiments conducted with both the Surrogate Model and Target Model as BiGCN.}
    \label{fig:3}
    
\end{figure}

\subsubsection{Hyperparameter \( \alpha_1 \)}

\(\alpha_1\) affects the weight of the supervised loss for the augmented data. In this experiment, we performed a sensitivity analysis on the hyperparameter \( \alpha_1 \). Specifically, we adjusted the value of \( \alpha_1 \) and compared the model's performance under different settings. The experiments were conducted using BiGCN surrogate and target models for ablation studies. Figure \ref{fig:3} shows the trend of changes in ACC. and AUA. values of the model on the Twitter and Weibo datasets under different \( \alpha_1 \) values.

 As shown in the experimental results, adjusting \(\alpha_1\) has a certain impact on the performance of SINCon, both on the Twitter and Weibo datasets. The ACC. in Normal and Normal+SINCon are similar, but when \(\alpha_1\) is too large or too small, Normal+SINCon performance slightly decreases. Similarly, for AUA., the performance of Normal+SINCon drops when \(\alpha_1\) is extreme.
 This is because the model tends to overfit the original MPT during the training process.

\subsubsection{Hyperparameter \( \alpha_2 \)}

\begin{figure}
    \centering
    \includegraphics[width=0.9\linewidth]{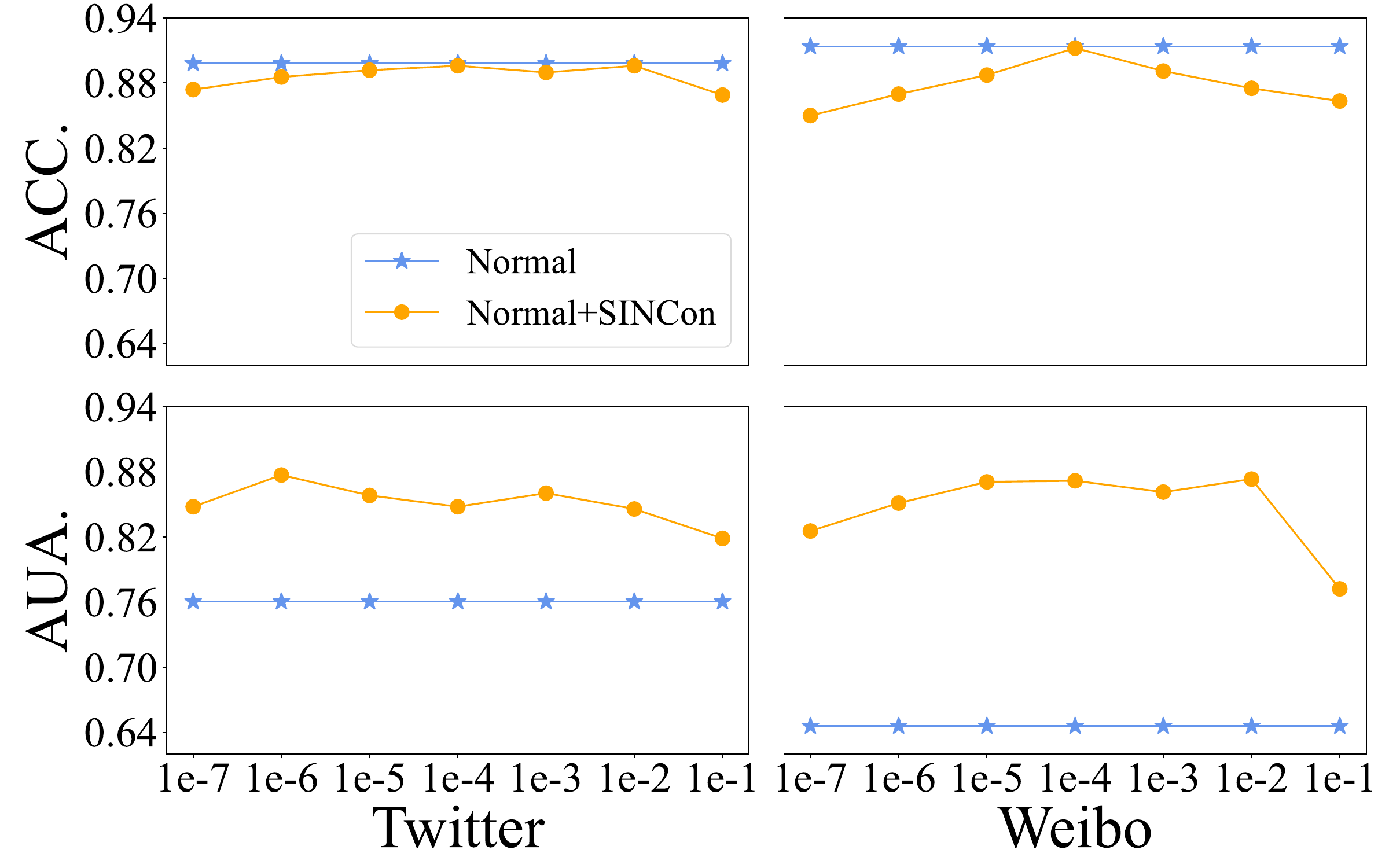}
    \vspace{-0.3cm}
    \caption{Sensitivity analysis of hyperparameters \(\alpha_2\). Experiments conducted with both the Surrogate Model and Target Model as BiGCN.}
    \label{fig:4}
    
\end{figure}

This weight affects the result of \themodel by affecting the weight of contrastive loss in the total loss. 
In this experiment, we conducted a sensitivity analysis of the hyperparameter $\alpha_2$ to assess its impact on model performance. The experiment used BiGCN-based surrogate and target models, and by adjusting the value of $\alpha_2$, we observed the variations in model performance (ACC. and AUA.) on the Twitter and Weibo datasets. From the experimental results shown in Figure \ref{fig:4}, it is evident that $\alpha_2$ has a noticeable impact on model performance. $\alpha_2$ affects the weight of the $\mathcal{L}_{\text{SINCon}}$ in the total loss function. First, regarding ACC, both excessively large and small values of $\alpha_2$ result in a certain degree of performance degradation. Compared to ACC., $\alpha_2$ has a more significant effect on AUA.. The results indicate that by balancing the influence of nodes in the MPTs, SINCon greatly enhances robustness against LLM-based message injection attacks.

%% file: conclusion.tex
\section{Conclusion}

In this paper, we proposed \themodel, a defense mechanism for enhancing the robustness of MPT-based rumor detection models against adversarial message injection attacks. By leveraging contrastive learning, \themodel ensures that both important and unimportant nodes exert more uniform influence on the model’s predictions, effectively mitigating the impact of localized perturbations caused by malicious message injections. Through extensive experiments on Twitter and Weibo datasets, we demonstrated that \themodel significantly improves the model's resilience to LLM-driven attacks while maintaining high classification accuracy on clean data. 

%% file: limitation.tex
\section{Limitations}
SINCon significantly enhances the performance of rumor detection models against LLM-driven message injection attacks, though at the cost of a slight decline in performance on clean data(an average of 1.48\%). Future research could further explore how to optimize data augmentation strategies and loss function design, aiming to improve the model's defensive robustness while maintaining high accuracy on clean data.
Moreover, this paper primarily focuses on LLM-driven malicious message injection attacks. However, in real-world environments, the methods of attack are becoming increasingly diverse. Future research should further examine the performance of SINCon against other types of attacks and explore more generalizable defense mechanisms.

%% file: appendix.tex
\section{Appendix}
\subsection{system prompt \( p \)}
\label{A.1}
\begin{mybox}
\textbf{Instruction}: Your mission is to construct a sentence that bears the least semantic similarity to the user’s inputs while maintaining a similar overarching topic. Cosine
similarity will be used to evaluate the dissimilarity.
\end{mybox}

\subsection{iterative prompting \( p' \)}
\label{A.2}
\begin{mybox}
\textbf{Instruction}: The similarity between the generated sentence and the input sentence is \(\{\text{sim}_{\text{root}}(G)\}\). 

Please generate a new sentence.
\end{mybox}

\subsection{Datasets}
\begin{table}[ht]
\centering
\resizebox{0.5\textwidth}{!}{
    \begin{tabular}{lcc}
    \toprule
    \textbf{Statistics} & \textbf{Twitter} & \textbf{Weibo} \\
    \midrule
    Users\#          & 491229 &2746818    \\
    Posts\#          & 1101985 &3805656    \\
    MPTs\#     & 992  & 4664 \\
    Rumors\#     & 498  & 2313    \\
    Non-Rumors\#          & 494  & 2351    \\
    Avg. time length/MPT  & 1582.6 Hours   & 2460.7 Hours   \\
    Avg \# of posts/MPT  & 1111   & 816   \\
    Max \# of posts/MPT   & 62827  & 59318 \\
    Min \# of posts/MPT   & 10  & 10 \\
    Language              & English & Chinese \\
    \bottomrule
    \end{tabular}
    }
\caption{ Statistics of the datasets.}
\label{tab:3}%
\end{table}